\documentclass{iau}
\usepackage{graphicx}
\def\sss{\scriptscriptstyle}
\def\U{{\sss \!U}}
\def\L{{\sss \!L}}
\def\K{{\sss \!K}}

\def\P{{\sss \!P}}
\def\nur{\nu_\mathrm{r}}
\def\nuv{\nu_\theta}
\def\nuL{\nu_\L}
\def\nuU{\nu_\U}
\def\nuK{\nu_\K}

\title[QPO models and GRS~1915$+$105] %% give here short title
{Confronting the Models of 3\,:\,2 QPOs with the Evidence of Near Extreme Kerr Black Hole}

\author[A. Kotrlov\'a et al.]   %% give here short author list %%
{Andrea Kotrlov\'a, Gabriel T\"{o}r\"{o}k, Eva \v{S}r\'{a}mkov\'{a} \and Zden\v{e}k Stuchl\'{\i}k}

\affiliation{Institute of Physics, Faculty of Philosophy and Science, Silesian University in Opava,\\
Bezru\v{c}ovo n\'{a}m. 13, CZ-74601 Opava, Czech Republic\\ email: {\tt andrea.kotrlova@fpf.slu.cz} \\[\affilskip]
}

\pubyear{2012}
\volume{290}  %% insert here IAU Symposium No.
\jname{Feeding compact objects: Accretion on all scales}
\editors{C.M. Zhang, T. Belloni, M. M\'endez \& S.N. Zhang, eds.}

\begin{document}

\maketitle
%--------------------------------------------------------
%--------------------------------------------------------
%												Abstract
%--------------------------------------------------------
%--------------------------------------------------------
\begin{abstract}

The black hole mass and spin estimates assuming various specific models of the 3\,:\,2 high frequency quasi-periodic oscillations (HF QPOs) have been carried out in \cite[T{\"{o}}r{\"{o}}k~{et~al.}~(2005,~2011)]{tor-etal:2005,Tor-Kot-Sra-Stu:2011:}. Here we briefly summarize some current points. Spectral fitting of the spin $a\equiv cJ/GM^2$ in the microquasar GRS~1915$+$105 reveals that this system can contain a near extreme rotating black hole (\cite[e.g., McClintock et al., 2011]{mcc-etal:2011}).  Confirming the high value of the spin would have significant consequences for the theory of the HF QPOs. The estimate of $a>0.9$ is almost inconsistent with the relativistic precession (RP), tidal disruption (TD), and the warped disc (WD) model. The epicyclic resonance (Ep) and discoseismic models assuming the c- and g- modes are instead favoured. However, consideration of all three microquasars that display the 3\,:\,2 HF QPOs  leads to a serious puzzle because the differences in the individual spins, such as $a=0.9$ compared to $a=0.7$, represent a generic problem almost for any unified orbital 3:2 QPO model.
\keywords{accretion, accretion disks; X-rays: binaries; black hole physics}
\end{abstract}

\firstsection % if your document starts with a section,
%													The spin
%--------------------------------------------------------
%--------------------------------------------------------
\section{The spin implied by individual models}

Assuming the Kerr geometry, the Keplerian and epicyclic orbital frequencies ($\nuK,~\nur$ and $\nuv$) for a given radius depend only on mass $M$ and spin $a$ of the black hole. It is therefore possible to infer the black hole spin or mass from the observed 3\,:\,2 frequencies and concrete orbital QPO models.

The 3\,:\,2 QPO frequencies in GRS~1915$+$105 are given by
%-------------------------------------------------------------------------------
\begin{equation}
\label{equation:3:2}
    \nuU=168\pm3\,\mathrm{Hz}\,,\quad\nuL=113\pm5\,\mathrm{Hz}\,.
\end{equation}
%-------------------------------------------------------------------------------
Assuming relation (\ref{equation:3:2}) and the well known formulae for the orbital frequencies, we calculate the implied mass-spin functions for the models associating the 3\,:\,2 QPOs with common radii by means of the definition relations given in Table~\ref{table1}. Following~\cite{tor-etal:2005} and taking into account the estimated range of the mass of GRS~1915$+$105,
%-------------------------------------------------------------------------------
\begin{equation}
\label{equation:mass}
    10\,\mathrm{\mathrm{M}}_{\odot} \le M \le 18\,\mathrm{\mathrm{M}}_{\odot},
\end{equation}
%-------------------------------------------------------------------------------
we infer the expected ranges of the spin. The results are presented in Table~\ref{table1}.

%--------------------------------------------------------
%--------------------------------------------------------
%													Table
%--------------------------------------------------------
%--------------------------------------------------------
\begin{table}
  \begin{center}
  \caption{Frequency relations corresponding to individual QPO models examined by \cite{Tor-Kot-Sra-Stu:2011:} and the resulting ranges of spin implied by the 3\,:\,2 QPOs in GRS 1915+105.}
  \label{table1}
\vspace{1ex}
\tabcolsep=6pt
 {\scriptsize
\begin{tabular}{|l|l|l|c|r|}
    \hline
     \textbf{Model} & \multicolumn{2}{|c|}{\textbf{Frequency relations}} & $\mathbf{\nuK/\nur}$\,\ \textbf{or~}\,$^*\mathbf{\nuv/\nu_\mathrm{r}}$&~~\textbf{a} $\sim$\\
    \hline
$\mathbf{RP}\phantom{1}$ & $\nuL =\nuK-\nur$ & $\nuU =\nuK$ & $3/1\phantom{^*}$ & $<0.55$\\
$\mathbf{TD}\phantom{1}$ & $\nuL=\nuK$ & $\nuU=\nuK+\nur$ & $2/1\phantom{^*}$ & $<0.45$\\
%\hline
$\mathbf{WD\phantom{1}}$ & $\nuL=2\left(\nuK-\nur\right)$ & $\nuU=2\nuK-\nur$ & $2/1\phantom{^*}$& $<0.45$\\
$\mathbf{Ep\phantom{1}}$ & $\nuL=\nur$ & $\nuU=\nuv$ & $3/2{^*}$& $0.65 - 1$\\
$\mathbf{Kep\phantom{1}}$ & $\nuL=\nur$ & $\nuU=\nuK$ & $3/2\phantom{^*}$ & $0.70 - 1$\\
$\mathbf{RP1\phantom{1}}$ &  $\nuL=\nuK-\nur$ &  $\nuU= \nuv$ &--$\phantom{^*}$& $<0.80$\\
$\mathbf{RP2\phantom{1}}$ &  $\nuL=\nuK-\nur$ &  $\nuU= 2\nuK-\nuv$ & --$\phantom{^*}$& $<0.45$\\
\hline
\end{tabular}
}
\end{center}
\vspace{1mm}
 \scriptsize{
 {\it Note:} The middle column indicates the ratio of the epicyclic frequencies determining the radii corresponding to the observed 3\,:\,2 ratio. The indicated ranges of spin also represent total spin ranges for the whole group of the three microquasars.}
\end{table}
%--------------------------------------------------------
%--------------------------------------------------------
%			  							End of table
%--------------------------------------------------------
%--------------------------------------------------------

The above considered models assume that both of the observed 3\,:\,2 frequencies are produced by the same mechanism and excited at a certain (common) preferred radius. For the discoseismic modes the individual observed QPOs correspond to different modes located at different radii. The frequencies of these modes depend on the black hole spin and the speed of sound in the accreted gas, and scale as $1/M$. The mass ranges implied by combinations of the fundamental discoseismic modes overlap with those observationally determined only for the model relating the 3\,:\,2 QPOs to the c-mode (corrugation vertically incompressible waves near the inner edge of the disk) and g-mode (inertial-gravity waves that occur at the radius where $\nur$ reaches its maximum value) provided that $0.90 \leq a \leq0.94$. Details and references are given in \cite{Tor-Kot-Sra-Stu:2011:}.

%--------------------------------------------------------
%--------------------------------------------------------
%													Conclusions
%--------------------------------------------------------
%--------------------------------------------------------
\section{Conclusions}

The internal (epicyclic) resonance and the discoseismic model (dealing with c- and g- modes) are favoured in the case of GRS~1915$+$105 provided that $a>0.9$. On the contrary, the TD, WD, RP, and RP2 models are disfavoured. This statement was inferred assuming that $\nu_\mathrm{K}$, $\nu_{r}$, and $\nu_{\theta}$ were the exact geodesic frequencies. Analysis including the influence of non-geodesic effects would require a very detailed study. A rough estimate of their possible relevance can be done assuming the relative non-geodesic correction $\Delta\nu$ (\cite[T{\"{o}}r{\"{o}}k {et~al.}, 2011]{Tor-Kot-Sra-Stu:2011:}),
which is needed to match the observations of GRS~1915$+$105 with a given model for a certain spin. For $a\in(0.9,1)$ and the RP model, it changes from $-40\%$ to $-60\%$. The same is roughly true for the TD and WD models, while for the RP2 model the required correction is even higher. Thus, the above result is justified, except when very large non-geodesic corrections are taken into account. Only the RP1 model can survive with corrections of $|\Delta\nu|$ up to $\sim20\%$, but the present physical interpretation of this model is unclear (see \cite[T{\"{o}}r{\"{o}}k {et~al.}, 2011]{Tor-Kot-Sra-Stu:2011:} for references).

\cite{tor-etal:2005} pointed out that since the 3\,:\,2 QPO frequencies in microquasars scale roughly as $\nu_{\mathrm{U}}\doteq 2.8\,(M/\mathrm{M}_{\odot})^{-1}\,\mathrm{kHz}$, %\cite{mcc-rem:2003},
their spins implied by a given resonance model should not much vary among them. If very different spins  in GRS~1915$+$105, GRO~J1655$-$40 and XTE~J1550$-$564  were confirmed, the difficulty of matching all the observed 3\,:\,2 frequencies would clearly be rather generic for most of the orbital QPO models.

Because of the generic $1/M$ scaling, the above difficulty also arises for a unified 3\,:\,2 QPO model assuming fundamental discoseismic modes.
%--------------------------------------------------------
%--------------------------------------------------------
%													Acknowledgements
%--------------------------------------------------------
%--------------------------------------------------------
\\\\
\textbf{Acknowledgements.}
The authors acknowledge the research grant GA\v{C}R~209/12/P740 and the project CZ.1.07/2.3.00/20.0071 -- ``Synergy'' supporting international collaboration of the Institute of Physics at SU Opava.

%--------------------------------------------------------
%--------------------------------------------------------
%													References
%--------------------------------------------------------
%--------------------------------------------------------

\end{document}